\documentclass[lettersize,journal]{IEEEtran}
\usepackage{amsmath,amsfonts}
\usepackage{algorithmic}
\usepackage{algorithm}
\usepackage{array}
\usepackage{textcomp}
\usepackage{stfloats}
\usepackage{url}
\usepackage{verbatim}
\usepackage{graphicx}
\usepackage{subcaption}
\usepackage{caption}
\usepackage{cite}
\usepackage{bm}
\usepackage{amsthm,amsmath,amssymb}
\usepackage{mathrsfs}
\usepackage{soul,color,xcolor}
\usepackage{graphicx}
\usepackage{epstopdf}
\usepackage{stfloats}

\allowdisplaybreaks[4]
\def\BibTeX{{\rm B\kern-.05em{\sc i\kern-.025em b}\kern-.08em
    T\kern-.1667em\lower.7ex\hbox{E}\kern-.125emX}}
\usepackage{balance}
\begin{document}
\title{Pinching Antenna-Aided Wireless Powered Communication Networks}
\author{Yixuan Li, Hongbo Xu, Ming Zeng, and Yuanwei Liu, \IEEEmembership{Fellow,~IEEE}
\thanks{\textit{Corresponding author: Hongbo Xu.}}
\thanks{Yixuan Li and Hongbo Xu are with the Department of Electronics and Information Engineering, Central China Normal University, Wuhan 430079, China (e-mail: yixuanli@mails.ccnu.edu.cn; xuhb@ccnu.edu.cn ).

Ming Zeng is with the Department of Electrical Engineering and Computer Engineering, Universite Laval, Quebec, QC G1V 0A6, Canada (e-mail: ming.zeng@gel.ulaval.ca).

Yuanwei Liu is with the Department of Electrical and Electronic Engineering, The University of Hong Kong, Hong Kong, China (e-mail: yuanwei@hku.hk).
}}

\maketitle

\begin{abstract}
In this letter, we investigate a novel pinching antenna (PA)-aided wireless powered communication network (WPCN), in which multiple PAs are activated along a waveguide to establish robust line-of-sight links with multiple devices. Both time division multiple access (TDMA) and non-orthogonal multiple access (NOMA) protocols are considered in the PA-WPCN. Moreover, some practical considerations, including a proportional power model for the PAs, a waveguide transmission loss model, and a nonlinear energy harvesting model, are incorporated into the PA-WPCN. Furthermore, we formulate a sum-rate maximization problem by jointly optimizing resource allocation and PAs position. To address the challenging problem of the PAs position optimization, we propose a high-performance element-wise (EW) algorithm and a low-complexity stochastic parameter differential evolution (SPDE) algorithm. Numerical results validate the remarkable performance of the proposed PA-WPCN and the effectiveness of our algorithms, indicating that optimal performance is attained when the PA power distribution ratio of approximately $\bf 0.55$$\bm -$$\bf 0.6$.
\end{abstract}


\section{Introduction}
The wireless powered communication networks (WPCNs) represent one of the novel wireless communication architectures that effectively address the challenges of power-limited Internet of Things (IoT) networks \cite{7462480}. In WPCNs, a hybrid access point (HAP) first transmits energy signals during the wireless power transfer (WPT) phase to energize power-limited IoT devices, and then receives information signals from the devices via the harvested energy during the wireless information transmission (WIT) phase, thereby enabling self-sustainable communication of the devices. However, the path loss of wireless communications exhibits significant distance dependency, and the combined effects of small-scale fading and large-scale fading push the received signal power to extremely low levels, especially for IoT devices located at cell edges. This results in inadequate energy transfer in downlink WPT and poor information transmission in uplink WIT, thereby degrading the system throughput.

To address these challenges, emerging solutions such as reconfigurable intelligent surfaces (RISs) \cite{8741198}, movable antennas \cite{10286328}, and fluid antennas \cite{9264694} have been proposed to enhance signal transmission by actively controlling the wireless propagation environment and reconfiguring channel characteristics. {Although these techniques are capable of alleviating small-scale fading \cite{10634973,10086660,9403371}, they remain insufficient for combating large-scale fading effects.} The pinching-antenna system (PASS) has been identified as a potential enabler for mitigating severe path loss and overcoming line-of-sight (LoS) link obstructions \cite{liu2025pinching}. By carefully configuring the activation points of pinching antennas (PAs) along the waveguide, stable LoS connections between the PAs and IoT devices can be reliably maintained. Existing works have fully demonstrated the significant performance improvements enabled by the PASS in wireless communications \cite{10909665,wang2025modeling,zeng2025energy,10896748,10976621,li2025,papanikolaou2025resolving}. In \cite{10909665} and \cite{10896748}, the authors demonstrated that PASS achieves superior performance compared to traditional architectures in terms of both uplink and downlink transmission efficiency. The authors of \cite{10976621} demonstrated that the PASS outperforms conventional systems in both reliability and data rate via performance analysis. To date, research efforts focusing on the integrating of PASS with wireless power transfer (WPT) remain scarce. In \cite{li2025}, PASS was shown to significantly improve the efficiency of simultaneous wireless information and power transfer (SWIPT) systems. In \cite{papanikolaou2025resolving}, the authors demonstrated that PASS can effectively mitigate the doubly-near-far problem commonly encountered in WPCNs.

In this letter, we investigate a novel pinching antenna-aided WPCN under time division multiple access (TDMA) and non-orthogonal multiple access (NOMA) protocols, where the proportional power model for the PAs, the waveguide transmission loss model, and the nonlinear energy harvesting model are considered. We jointly optimize resource allocation and PAs position to maximize the sum achievable rate of the WP-WPCN. To address the PAs position optimization problem, we propose a high-performance element-wise (EW) algorithm and a low-complexity stochastic parameter differential evolution (SPDE) algorithm. Simulation results demonstrate the superior performance of the PA-WPCN compared to conventional fixed-antenna systems and verify that the optimal PA power distribution ratio is approximately $0.55$$-$$0.6$. 

\section{System Model and Problem Formulation}
\begin{figure}[!t]
\centering
\includegraphics[width=2.2in]{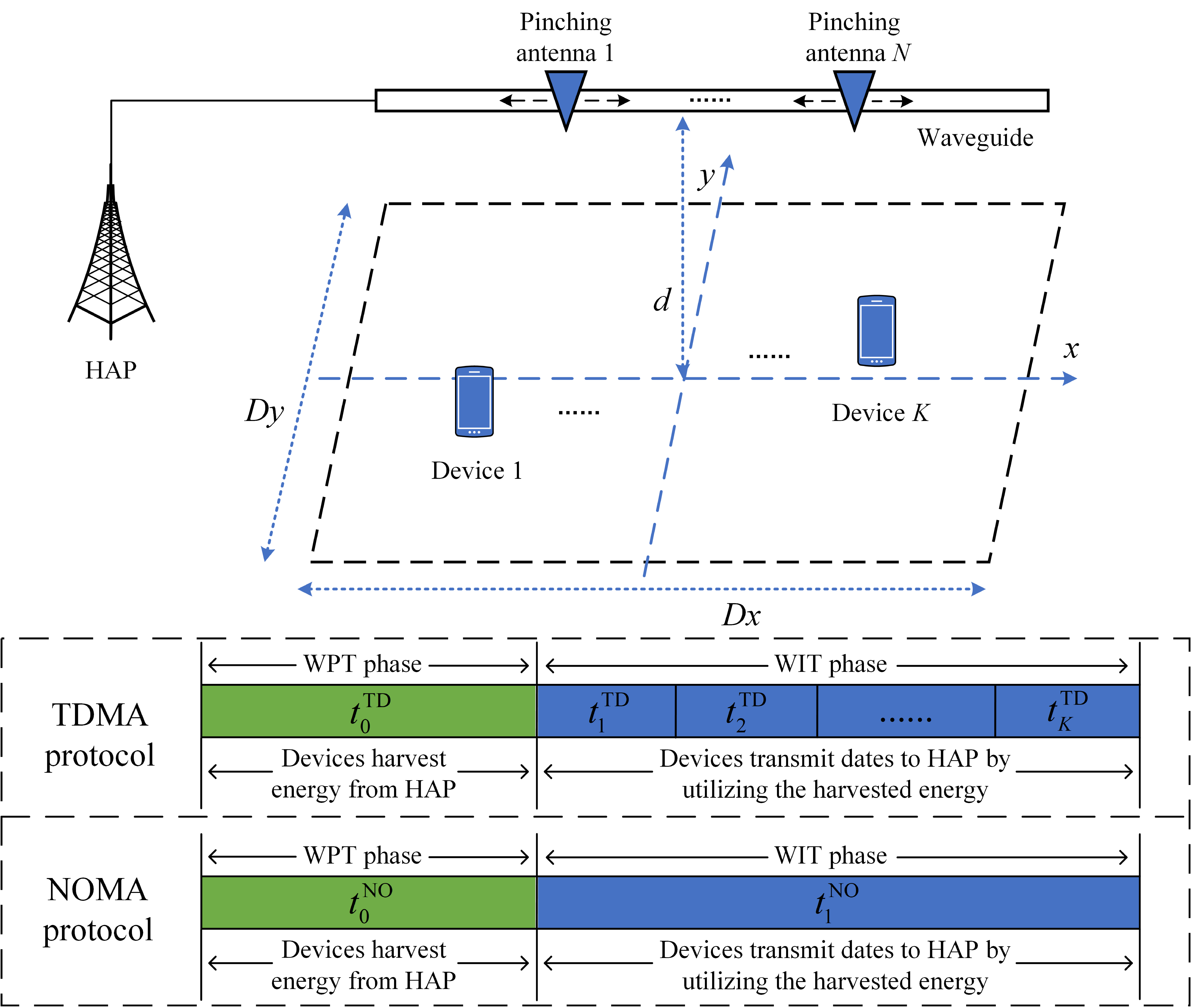}
\caption{System model and operation protocols of the proposed {PA-WPCN}.}
\label{system_model}
\end{figure}
Fig. \ref{system_model} illustrates the system model of the proposed PA-WPCN, where $N$ PAs are activated on a waveguide connected with the hybrid access point (HAP) to serve $K$ IoT devices. The sets of the PAs and the devices are denoted as ${\cal N} = \left\{1,...,N\right\}$ and ${\cal K} = \left\{1,...,K\right\}$, respectively. The coordinates of the $n$-th PA and the $k$-th device are expressed as ${\bf r}_n^{\rm PA} = \left[x_n^{\rm PA},0,d\right]^T$ and ${\bf r}_k^{\rm D} = \left[x_k^{\rm D},y_k^{\rm D},0\right]^T$, respectively, where $\forall n \in {\cal N}$ and $\forall k \in {\cal K}$. To prevent signal interference between energy and information transmissions on the same waveguide, the system employs a Harvest-Then-Transmit (HTT) protocol to facilitate downlink WPT and uplink WIT. Specifically, the HAP sends energy signals through the PAs to the devices in WPT phase. Subsequently, the devices utilize the harvested energy to transmit information signals back to the HAP via the PAs in WIT phase, thereby enabling self-sustaining communication. To fully leverage the potential of the PA-WPCN, the system adopts both TDMA and NOMA protocols. The specific operations of the two protocols are illustrated in the lower portion of Fig. \ref{system_model}. The time slot allocated for WPT is denoted as $t_0^{\rm TD/NO}$. Under the TDMA protocol, $t_k^{\rm TD}$ represents the time slot allocated to the $k$-th device. Under the NOMA protocol, $t_1^{\rm NO}$ denotes the time slot allocated to all devices simultaneously. 

In this work, we adopt the proportional power model \cite{wang2025modeling} and incorporate waveguide transmission loss model, which are more suitable for practical scenarios{\footnote{In practice, hardware impairments may cause deviations in the proportional power model. Since PAs are passive and lack amplitude or phase control, calibration must rely on geometric methods such as position estimation or deployment-aware optimization, which merit further study.}}. The channels between the PAs and the devices are assumed near-field channels owing to the large antenna aperture of the PASS \cite{10896748}. Therefore, the channel between the $n$-th PA and the $k$-th device is given by
\begin{flalign}
{{h}_{n,k}} = {{\frac{{{\eta ^{\frac{1}{2}}}{e^{ - j\frac{{2\pi }}{\lambda }\left\| {{\bf r} _k^{{\rm{D}}} - {\bf r} _n^{{\rm{PA}}}} \right\|}  }}}{{\left\| {{\bf r} _k^{{\rm{D}}} - {\bf r} _n^{{\rm{PA}}}} \right\|}}} \cdot {\beta_n^{\frac{1}{2}}}{\varsigma_n^{\frac{1}{2}}} {e^{ - j\frac{{2\pi }}{{{\lambda _g}}}\left\| {{\bf r} _0^{\rm PA} - {\bf r} _n^{\rm PA}} \right\|}}},
\end{flalign}
where $\lambda$ and $\lambda _g = \frac{\lambda}{n_{\rm neff}}$ denote the wavelengths in the free-space and the waveguide, respectively. $\eta ^{\frac{1}{2}} = \frac{c}{{4\pi {f_c}}}$ and $n_{\rm neff}$ are constants, where $c$ and $f_c$ represent the speed of light and frequency, respectively. The symbol ${\bf r}_0^{\rm PA}$ is used to represent the feed point coordinate. $\beta _n = \delta ^2 \left(1-{\delta ^2}\right)^{n-1}$ and $\varsigma_n = 10^{-\frac{\mu \left\|{\bf r}_0^{\rm PA} - {\bf r}_n^{\rm PA}\right\|}{10}}$ represent power distribution ratio and power loss coefficient at the $n$-th PA, respectively. Here, $\delta = \sin \left(\kappa L\right)$ is the power distribution factor for the PAs, where $\kappa$ and $L$ denote the coupling coefficient and coupling length between the PAs and the waveguide, respectively. $\mu$ denotes the power loss coefficient of the signal along the waveguide. For the convenience of subswquent calculations, the joint channel between the HAP and the $k$-th device is denoted as $h_k = \sum\nolimits_{n = 1}^N {h_{n,k}}$. 

Furthermore, a more realistic nonlinear energy harvesting (EH) model is adopted to characterize the practical scenarios, which can be represented as ${\Phi _k} = \frac{{{\Psi _k} - {Z_k}{\Omega _k}}}{{1 - {\Omega _k}}}$, where ${\Psi _k} = \frac{{{Z_k}}}{{1 + {e^{ - {a_k}\left( {{P_0}{{\left| {{h_k}} \right|}^2} - {b_k}} \right)}}}}$ and ${\Omega _k} = \frac{1}{{1 + {e^{{a_k}{b_k}}}}}$ \cite{7264986}. $Z_k$, $a_k$, and $b_k$ are the related hardware parameters in the nonlinear EH model for the $k$-th device. Thus, the harvested energy at the $k$-th device in {WPT} phase is $E_k^{\rm TD/NO} = t_0^{\rm TD/NO} {\Phi _k}$. In this work, we make a more practical assumption that the positions of the PAs remain fixed throughout the entire time period. 

\subsection{Problem Formulation for TDMA and NOMA}
In WIT phase, the devices send data to the HAP in sequence utilizing the harvested energy during WPT phase. The achievable rate at the $k$-th device under TDMA protocol is represented as
\begin{flalign}
R_k^{\rm TD} &= {t_k^{\rm TD}}{\log _2}\left( {1 + \frac{{{{p_k^{\rm TD}}{{\left| {{h_{k}}} \right|}^2}} }}{{{\sigma _k^{2}}}}} \right),
\end{flalign}
where $p_k^{\rm TD}$ and $\sigma _k^2$ are the transmit power and the noise power at the $k$-th device, respectively. Let ${\bf x} = \left\{x_1^{\rm PA},...,x_N^{\rm PA}\right\}$, ${\bf t}^{\rm TD} = \left\{t_1^{\rm TD},...,t_K^{\rm TD}\right\}$, and ${\bf p}^{\rm TD} = \left\{p_1^{\rm TD},...,p_N^{\rm TD}\right\}$. To maximize the achievable sum rate of the PA-WPCN, the optimization problem under the TDMA can be formulated as
\begin{subequations}\label{P1}
\begin{eqnarray}
&\mathop {\max }\limits_{{\bf x},{t_0^{\rm TD}},{\bf t}^{\rm TD},{\bf p}^{\rm TD}} &\sum\nolimits_{k = 1}^K {R_k^{\rm TD}} \label{P1-a}\\
&{\rm s.t.} 
&{t_k^{\rm TD}}\left({p_k^{\rm TD}}+{p_{c,k}}\right) \le {t_0^{\rm TD} {\Phi _k}},\forall k \in {\cal K}, \label{P1-c}\\
&&\sum\nolimits_{k = 0}^K {{t_k^{\rm TD}}}  \le 1, \label{P1-d}\\
&&p_k^{\rm TD}, t_0^{\rm TD}, t_k^{\rm TD} \ge 0, \forall k \in {\cal K}, \label{P1-e}\\
&&{x_n^{\rm PA} - x_{n-1}^{\rm PA}} \ge \Delta ,\forall n \in \left\{2,...,N\right\},~~~\label{P1-f}\\
&&x_n^{\rm PA} \in \left[ {0,{D_w}} \right],\forall n \in {\cal N}. \label{P1-g}
\end{eqnarray}
\end{subequations}

In WIT phase, the successive interference cancellation (SIC) technology is adopted when employing NOMA protocol in this system. {To promote user fairness, the IoT devices are arranged in descending order according to their effective channel gains, i.e., $\left| {{h_1}} \right| \ge  \cdots  \ge \left| {{h_K}} \right|$, and are subsequently decoded following this order using SIC.} Noting, the implementation of any SIC decoding sequence exerts no influence on the achievable sum rate in the uplink NOMA systems. In the following sections, we will demonstrate this conclusion. Thus, the achievable rate at the $k$-th device under NOMA protocol is denoted as
\begin{flalign}
R_k^{\rm NO} &= {t_1^{\rm NO}}{\log _2}\left( {1 + {\frac{{{p_k^{\rm NO}}{{\left| {{h_{k}}} \right|}^2}}}{{\sum\nolimits_{j = k + 1}^K {{p_j^{\rm NO}}{{\left| {{h_{j}}} \right|}^2}}  + {\sigma _k^2}}}} } \right),
\end{flalign}
where $p_k^{\rm NO}$ denotes the transmit power at the $k$-th device. Let ${\bf p}^{\rm NO} = \left\{p_1^{\rm NO},...,p_N^{\rm NO}\right\}$. Thus, the sum rate maximization problem under the NOMA protocol is formulated as
\begin{subequations}\label{P2}
\begin{eqnarray}
&\mathop {\max }\limits_{{\bf x},{t_0^{\rm NO}},{t_1^{\rm NO}},{\bf p}^{\rm NO}} &\sum\nolimits_{k = 1}^K {R_k^{\rm NO}} \label{P2-a}\\
&{\rm s.t.} 
&{t_1^{\rm NO}}\left({p_k^{\rm NO}}+p_{c,k}\right) \le {t_0^{\rm NO} {\Phi _k}},\forall k \in {\cal K},~~~ \label{P2-c}\\
&&t_0^{\rm NO} + t_1^{\rm NO} \le 1, \label{P2-d}\\
&&p_k^{\rm NO}, t_0^{\rm NO}, t_1^{\rm NO} \ge 0, \forall k \in {\cal K}, \label{P2-e}\\
&&\left({\rm \ref{P1-f}}\right),\left({\rm \ref{P1-g}}\right).
\end{eqnarray}
\end{subequations}

In problems (\ref{P1}) and (\ref{P2}), constraints (\ref{P1-c}) and (\ref{P2-c}) mean the transmitted energy of each IoT device cannot exceed its harvested energy, where $p_{c,k}$ denotes the internal circuit power consumption of the $k$-th device. Constraints (\ref{P1-d}), (\ref{P1-e}), (\ref{P2-d}), and (\ref{P2-e}) are the time slot and power allocation constraints, where the time of the entire period is denoted as $1~{\rm s}$. Constraints (\ref{P1-f}) and (\ref{P1-g}) are the position constraints of the PAs. Constraint (\ref{P1-f}) specifies that the spacing between adjacent PAs must not be less than $\Delta$ to avoid antenna coupling. Constraint (\ref{P1-g}) means all PAs must be activated on the waveguide.

\section{Proposed Algorithms}
It is worth noting that both problems (\ref{P1}) and (\ref{P2}) are challenging to solve due to coupling between variables ${\bf t}^{\rm TD/NO}$, ${\bf p}^{\rm TD/NO}$, and ${\bf x}$. To address them, each original problem is first decoupled into two subproblems: resource allocation subproblem and PAs position optimization subproblem.

\subsection{Optimal Resource Allocation for TDMA and NOMA}
Given $\bf x$, problem (\ref{P1}) can be simplified as
\begin{subequations}\label{P3}
\begin{eqnarray}
&\mathop {\max }\limits_{{t_0^{\rm TD}},{\bf t}^{\rm TD},{\bf p}^{\rm TD}} &\sum\nolimits_{k = 1}^K R_k^{\rm TD} \label{P3-a}\\
&{\rm s.t.} &\left({\rm \ref{P1-c}}\right),\left({\rm \ref{P1-d}}\right),\left({\rm \ref{P1-e}}\right).
\end{eqnarray}
\end{subequations}

The function (\ref{P3-a}) is non-concave and (\ref{P1-c}) is non-convex, thereby the above problem is still difficult to address. The optimal time slot and power allocation are achieved when constraint (\ref{P1-c}) holds with equality. That is, the optimal variables $t_k^{\rm TD}$ and $p_k^{\rm TD}$ must satisfy $t_k^{\rm TD} \left(p_k^{\rm TD}+p_{c,k}\right) = t_0^{\rm TD} {\Phi _k}$. Thus, problem (\ref{P3}) is recast as
\begin{subequations}\label{P3.1}
\begin{eqnarray}
&\mathop {\max }\limits_{{t_0^{\rm TD}},{\bf t}^{\rm TD}} &\sum\limits_{k = 1}^K {{t_k^{\rm TD}}{\log _2}\left( {A_k + \frac{{{{t_0^{\rm TD} {\Phi _k}}{{\left| {{h_{k}}} \right|}^2}} }}{{t_k^{\rm TD} {\sigma _k^{2}}}}} \right)} ~~~\label{P3.1-a}\\
&{\rm s.t.} &\left({\rm \ref{P1-d}}\right),\left({\rm \ref{P1-e}}\right),
\end{eqnarray}
\end{subequations}
where $A_k = 1 - {\frac{p_{c,k} \left|h_k\right|^2}{\sigma _k^2}}$.

Given $\bf x$, problem (\ref{P2}) is recast as
\begin{subequations}\label{P4}
\begin{eqnarray}
&\mathop {\max }\limits_{{t_0^{\rm NO}},{t_1^{\rm NO}},{\bf p}^{\rm NO}} &\sum\nolimits_{k = 1}^K {R_k^{\rm NO}} \label{P4-a}\\
&{\rm s.t.} &\left({\rm \ref{P2-c}}\right),\left({\rm \ref{P2-d}}\right),\left({\rm \ref{P2-e}}\right).
\end{eqnarray}
\end{subequations}

The above problem is also intractable, for similar reasons as the optimization problem of TDMA protocol, since $p_k^{\rm NO}$ and $t_1^{\rm NO}$ are mutually coupled. To address the issues, the sum rate of the PA-WPCN under NOMA protocol is transformed as
\begin{flalign}
&\sum\limits_{k = 1}^K {R_k^{\rm NO}} = {t_1^{\rm NO}}{\log _2}\left( {1 + \frac{{\sum\nolimits_{k = 1}^K {{p_k^{\rm NO}}{{\left| {{h_k}} \right|}^2}} }}{{{\sigma _k^{2}}}}} \right). \label{RNOMA}
\end{flalign}

Based on Eq. (\ref{RNOMA}), the sum rate of the PA-WPCN remains unaffected by the SIC decoding order. Similar to TDMA, the optimal $t_0^{\rm NO}$, $t_1^{\rm NO}$ and $p_k^{\rm NO}$ in NOMA protocol must also satisfy $t_1^{\rm NO} \left(p_k^{\rm NO} + p_{c,k}\right) = t_0^{\rm NO} {\Phi _k}$. Thus, problem (\ref{P4}) is rewritten as
\begin{subequations}\label{P4.1}
\begin{eqnarray}
&\mathop {\max }\limits_{{t_0^{\rm NO}},{t_1^{\rm NO}}} &{{{t_1^{\rm NO}}{{\log }_2}\left(S + {\frac{{\sum\nolimits_{k = 1}^K {{t_0^{\rm NO} {\Phi _k}}{{\left| {{h_k}} \right|}^2}} }}{{t_1^{\rm NO}}{\sigma _k^2}}} \right)}}~~~ \label{P4.1-a}\\
&{\rm s.t.}&\left({\rm \ref{P2-d}}\right),\left({\rm \ref{P2-e}}\right),
\end{eqnarray}
\end{subequations}
where $S = 1 - \sum\nolimits_{k=1}^{K}\frac{{p_{c,k} \left|h_k\right|^2}}{\sigma _k^2}$.

{It is noted that the arguments inside the ${\log _2}\left(  \cdot  \right)$ function in (\ref{P3.1-a}) and (\ref{P4.1-a}) are strictly positive, since the transmit powers are non-negative and the circuit power consumption $p_{c,k}$ is set within practical low-power ranges, ensuring system feasibility.} Problem (\ref{P3.1}) and (\ref{P4.1}) are both tractable convex optimization problem since the concave (\ref{P3.1-a}) and (\ref{P4.1-a}) and convex constraints. Therefore, the two optimization problems described above can be directly addressed by utilizing Karush-Kuhn-Tucker (KKT) conditions.

{\textit{Proposition 1:}} When $p_{c,k} = 0$, both TDMA and NOMA protocols can achieve the same sum-rate in the considered system. However, when $p_{c,k} > 0$, the sum-rate achieved by TDMA surpasses that of NOMA.

\begin{proof}
To apply Jensen's inequality, assuming $p_{c,k} = 0$, we redefine $\sum\nolimits_{k = 1}^K R_k^{\rm TDMA}$ to align with the required structure, thereby enabling the inequality to be reformulated as
\begin{flalign}
\sum\limits_{k = 1}^K R_k^{\rm TD} = \left( {\sum\limits_{k = 1}^K {{t_k^{\rm TD}}} } \right) \cdot \sum\limits_{k = 1}^K {{w_k}{{\log }_2}\left( {1 + \frac{{{t_0^{\rm TD} {\Phi _k}}{{\left| {{h_k}} \right|}^2}}}{{{w_k}\sum\limits_{i \in {\cal K}} {{t_i^{\rm TD}}}  \cdot {\sigma _k^2}}}} \right)},
\end{flalign}
where $w_k = t_k^{\rm TD} / \sum\nolimits_{i=1}^K {t_i^{\rm TD}}$. It should be satisfied by $\sum\nolimits_{k = 1}^K {w_k} = 1$. Based on the above, we have
\begin{align}
	&\sum_{k = 1}^K w_k \log_2\left( 1 + \frac{t_0^{\rm TD} {\Phi _k}|h_k|^2}{w_k\sum_{i=1}^{K} t_i^{\rm TD}\sigma_k^2}\right)\notag\\
	&~~~~~~~~~\leq \log_2\left(1 + \frac{\sum_{k = 1}^K t_0^{\rm TD} {\Phi _k}|h_k|^2}{\sum_{k = 1}^K t_k^{\rm TD}\sigma_k^2}\right).\label{aa}
\end{align}

Noted that the optimal time slot allocation should be satisfied by $t_0^{\rm TD} + \sum\nolimits_{k=1}^K {t_k^{\rm TD}} = 1$. By multiplying both sides of inequality (\ref{aa}) by $\sum\nolimits_{k \in {\cal{K}}} {t_k^{\rm TD}}$, it can be reformulated as
\begin{flalign}
&\sum\limits_{k=1}^K t_k^{\rm TD} {\log _2}\left( {1 + \frac{{{{t_0^{\rm TD} {\Phi _k}}{{\left| {{h_k}} \right|}^2}} }}{{{t_k^{\rm TD}} {\sigma _k^2}}}} \right) = \sum\limits_{k=1}^K {R_k^{\rm TD}} \notag\\
&~~~~~~\le \left( 1-t_0^{\rm TD} \right){\log _2}\left( {1 + \frac{{\sum\nolimits_{k = 1}^K {{t_0^{\rm TD} {\Phi _k}}{{\left| {{h_k}} \right|}^2}} }}{{\left(1 - t_0^{\rm TD} \right){\sigma _k^2}}}} \right),\label{bb}
\end{flalign}
where equality holds if and only if ${{{{t_0^{\rm TD} {\Phi _1}}{{\left| {{h_1}} \right|}^2}}}}/{{t_1{\sigma ^2}}} = ... = {{{{t_0^{\rm TD} {\Phi _K}}{{\left| {{h_K}} \right|}^2}}}}/{{t_K{\sigma ^2}}}$. 

We can observe that when $p_{c,k} = 0$, the TDMA and NOMA protocols share an identical sum-rate expression. Therefore, it can be concluded that under the fixed ${\bf x}$, TDMA and NOMA protocols achieve equal sum-rate performance. When $p_{c,k} > 0$, the NOMA protocol leads to higher total internal circuit power consumption across IoT devices compared to the TDMA protocol. Consequently, the TDMA protocol achieves a higher sum-rate than the NOMA protocol under these conditions. This completes the proof.
\end{proof}

%
%

\subsection{Position Optimization for PAs}
Given ${\bf t}^{\rm TD/NO}$ and ${\bf p}^{\rm TD/NO}$, problems (\ref{P1}) and (\ref{P2}) can be recast as
\begin{subequations}\label{P5}
\begin{eqnarray}
&\mathop {\max }\limits_{{\bf x}} &\sum\nolimits_{k = 1}^K {R_k^{{\rm TD}/{\rm NO}}} \label{P5-a}\\
&{\rm s.t.} 
&\left({\rm \ref{P1-f}}\right),\left({\rm \ref{P1-g}}\right).
\end{eqnarray}
\end{subequations}

We can find that problem (\ref{P5}) is rendered intractable to optimize. Notably, the PAs activation position optimization problem in PASS has numerous local optima with substantial differences among them, primarily due to the extensive feasible region of these positions \cite{wang2025modeling}. This characteristic renders traditional algorithms ineffective for PASS, such as gradient-descent algorithm. Therefore, we propose a two-step solution approach: first utilizing the EW algorithm to solve for $x_n^{\rm PA}$ to avoid being trapped in poor local solutions, and then employing a gradient ascent-based algorithm to obtain a near-optimal solution for $x_n^{\rm PA}$. 

\subsubsection{Element-wise algorithm}
We develop an element-wise algorithm to address $\bf x$. Specifically, problem (\ref{P5}) can be decomposed into $N$ subproblems with respect to the $N$ PAs position, respectively. Thus, the $n$-th subproblem for (\ref{P5}) is recast as
\begin{subequations}\label{P5.1}
\begin{eqnarray}
&\mathop {\max }\limits_{{x_n^{\rm PA}}} &\sum\nolimits_{k = 1}^K {R_k^{{\rm TD}/{\rm NO}}} \label{P5.1-a}\\
&{\rm s.t.} &\left({\rm \ref{P1-f}}\right),\left({\rm \ref{P1-g}}\right).
\end{eqnarray}
\end{subequations}

The positions of the PAs can be determined sequentially using a one-dimensional search approach. Specifically, starting from the first PA to the $N$-th PA, we iteratively search for the best position within the feasible region along the waveguide until convergence. The search step size $D$ can be flexibly adjusted to balance the requirements of search accuracy and computational complexity.

\subsubsection{Stochastic Parameter Differential Evolution Algorithm}
We proposed a SPDE algorithm to optimize the PAs position, leveraging its efficient global search capability and rapid convergence characteristics. The details are presented as follows.

\textit{{a. Encoding scheme and initialization:}}
Each candidate solution is encoded as a vector ${\bf x}$, representing the optimization variables $x_n^{\rm PA}$, $\forall n \in {\cal N}$, satisfying all constraints. The initial population $\left\{{\bf x}_i\right\}_{i=1}^{Q}$ is randomly generated within the feasible region defined by the constraints, where $Q$ denotes the population size.

\textit{{b. Mutation, crossover, and selection:}}
At iteration $g$, mutation is performed by generating mutant vectors ${\bf v}_i^g$ as
\begin{flalign}
	{\bf v}_i^g = {\bf x}_{r1}^g + F \times \left({\bf x}_{r2}^g-{\bf x}_{r3}^g\right), r1 \ne r2 \ne r3 \ne i,
\end{flalign}
where indices $r1$, $r2$, $r3$ are distinct random integers selected from $\left\{1,...,Q\right\}$, and $F = {\rm rand}\left(1\right)$ is a scaling factor randomly adjusted during the evolution process. Randomized factor settings enhance the algorithm's global search capability, thereby reducing the likelihood of becoming trapped in inferior local optima.

A trial vector ${\bf u}_i^g$ is generated by combining the mutant vector ${\bf v}_i^g$ and target vector ${\bf x}_i^g$ using binomial crossover as
\begin{flalign}
	u_{i,j}^g = \left\{ {\begin{array}{*{20}{c}}
			{v_{i,j}^g,}&{{\rm{if }}~{{\rm{rand}}_j} \le CR ~{\rm{ or }}~j = {j_{\rm rand},}}\\
			{x_{i,j}^g,}&{{\rm{otherwise.}}}
	\end{array}} \right.
\end{flalign}
Here, $CR = {\rm rand}\left(1\right)$ is the crossover probability randomly tuned by SPDE, ${\rm rand}_j$ is uniformly distributed within $\left[0,1\right]$, and $j_{\rm rand}$ ensures at least one parameter from ${\bf v}_i^g$.

Each trial vector ${\bf u}_i^g$ competes with the corresponding target vector ${\bf x}_i^g$, and the one with higher fitness (\ref{P5-a}) is selected for the next generation as
\begin{flalign}
	{\bf{x}}_i^{g + 1} = \left\{ {\begin{array}{*{20}{c}}
			{{\bf{u}}_i^g,}&{{\rm{if }}f\left( {{\bf{u}}_i^g} \right) \le f\left( {{\bf{x}}_i^g} \right),}\\
			{{\bf{x}}_i^g,}&{{\rm{otherwise.}}}
	\end{array}} \right.
\end{flalign}

{\textit{c. Constraint handling:}} {PA positions that violate the spacing constraint ${x_n^{\rm PA} - x_{n-1}^{\rm PA}} \ge \Delta$ are forcibly corrected by setting $x_n^{\rm PA} = x_{n-1}^{\rm PA} + \Delta$. Furthermore, PA positions falling outside the range $\left[0,D_w\right]$ are projected onto the nearest boundary.} 

Finally, the evolution continues until a predefined maximum number $g_{\max}$ is reached.

\subsection{Complexity and Convergence Analysis}
Problem (\ref{P1}) and (\ref{P2}) can be addressed via alternating optimization (AO) of the two subproblems. Since the objective function is non-decreasing and remains upper-bounded throughout the alternating optimization iterations, the algorithm's convergence is guaranteed. Moreover, the approximate computational complexities of the SPDE-based AO algorithm for the TDMA and NOMA protocols are characterized as ${\cal O}{\left( {L_{\rm DE}\left[K^4+g_{\max}QN\right]} \right)}$ and ${\cal O}{\left( {L_{\rm DE}\left[K^4+g_{\max}QN\right]} \right)}$, respectively. Furthermore, the approximate computational complexities of the EW-based AO algorithm for the TDMA and NOMA protocols are characterized as ${\cal O}{\left( {L_{\rm EW}\left[K^4+DN\right]} \right)}$ and ${\cal O}{\left( {L_{\rm EW}\left[K^4+DN\right]} \right)}$, respectively. Here, $L_{\rm DE}$, $L_{\rm EW}$ are the iteration number of the SPDE-based AO algorithm and the EW-based AO algorithm, respectively.

\section{Numerical Results}

\begin{figure}[!t]
\centering
\includegraphics[width=2.2in]{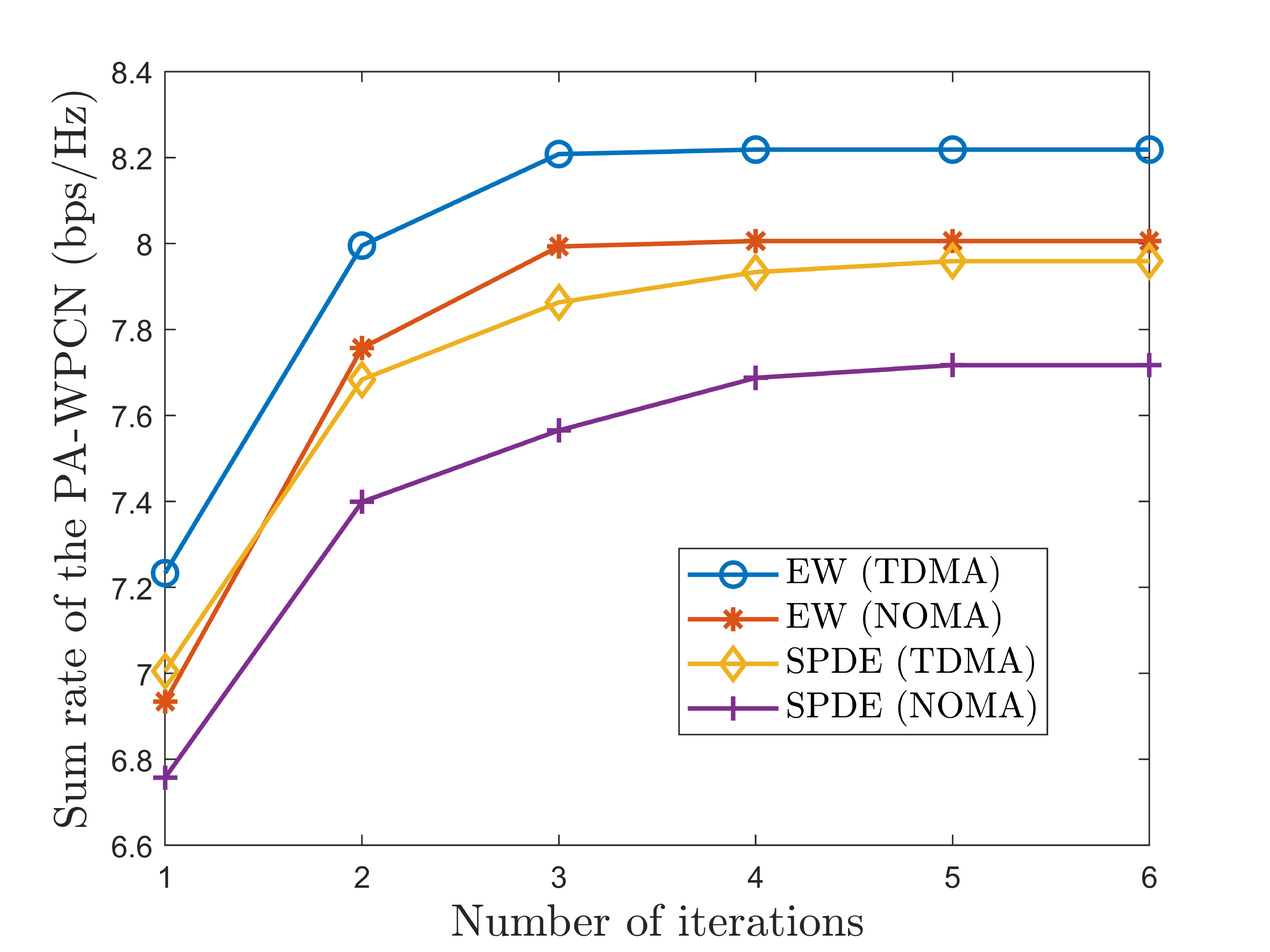}
\caption{Sum rate versus number of iterations.}
\label{iteration}
\end{figure}

This section demonstrates the performance advantages of {PA-WPCN} under the TDMA and NOMA protocols via numerical results. The parameter settings are as follows: $N = 6$, $K = 10$, $P_0 = 40~{\rm dBm}$, $\sigma _k^2 = -120~{\rm dBm}$, $f_c = 28~{\rm GHz}$, $n_{\rm neff} = 1.4$, $\delta = 0.6$, $\mu = 0.2$, $p_{c,k} = 10^{-7}~{\rm W}, \forall k$, $\Delta = \frac{\lambda}{2}$, $D = 2000$, $Q = 30$, $g_{\max} = 200$, $D_x = 10~{\rm m}$, $D_y = 6~{\rm m}$, $D_w = 10~{\rm m}$, $d = 3~{\rm m}$, $T = 1~{\rm s}$, {$a_k = 150$, $b_k = 0.014$, $Z_k = 0.024$}. We denote the proposed EW-based AO and SPDE-based AO algorithms for the PA-WPCN as `EW' and `SPDE', respectively. {An RIS-aided WPCN is denoted as `RIS'.} The fixed-antenna WPCN is denoted as `Conv'.

Fig. {\ref{iteration}} illustrates the convergence performance of the two proposed algorithms. It can be observed that both the proposed SPDE and EW algorithms exhibit rapid convergence characteristics. Moreover, the performance gap between the SPDE and the EW algorithms is only approximately 4\%. Additionally, the TDMA-based system achieves superior performance compared to the NOMA-based PA-WPCN, primarily due to its lower overall device circuit power consumption. 

\begin{figure}[!t]
	\centering
	\begin{subfigure}{0.49\linewidth}
		\centering
		\includegraphics[width=\linewidth]{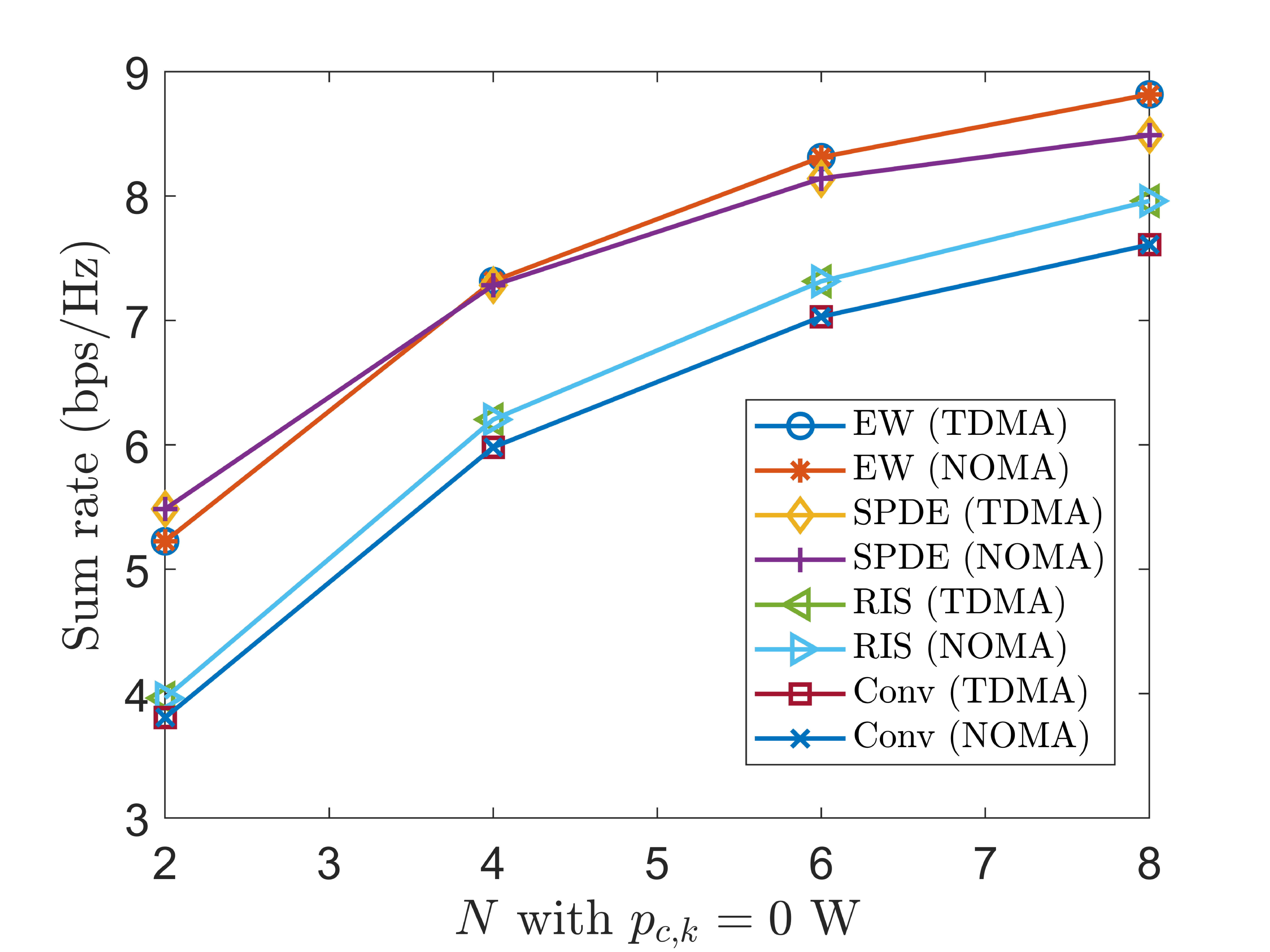}
		\subcaption{\centering $p_{c,k} = 0~{\rm W}$.}
		\label{Na}
	\end{subfigure}
	\hfill
	\begin{subfigure}{0.49\linewidth}
		\centering
		\includegraphics[width=\linewidth]{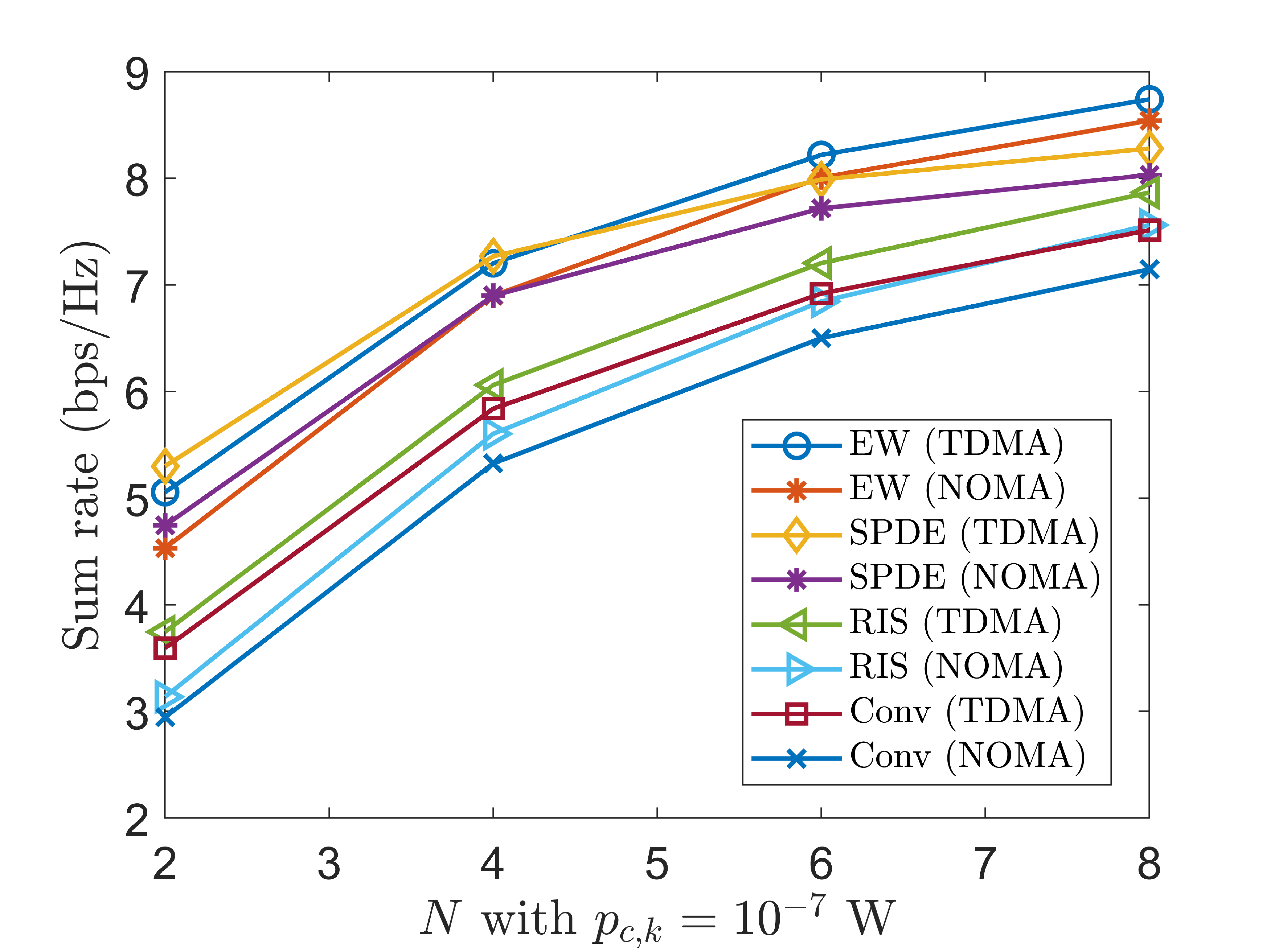}
		\subcaption{\centering $p_{c,k} = 10^{-7}~{\rm W}$.}
		\label{Nb}
	\end{subfigure}
	\caption{Sum rate versus $N$ with {different} $p_{c,k}$.}
	\label{N}
\end{figure} 

\begin{figure}[!t]
	\centering
	\begin{subfigure}{0.49\linewidth}
		\centering
		\includegraphics[width=\linewidth]{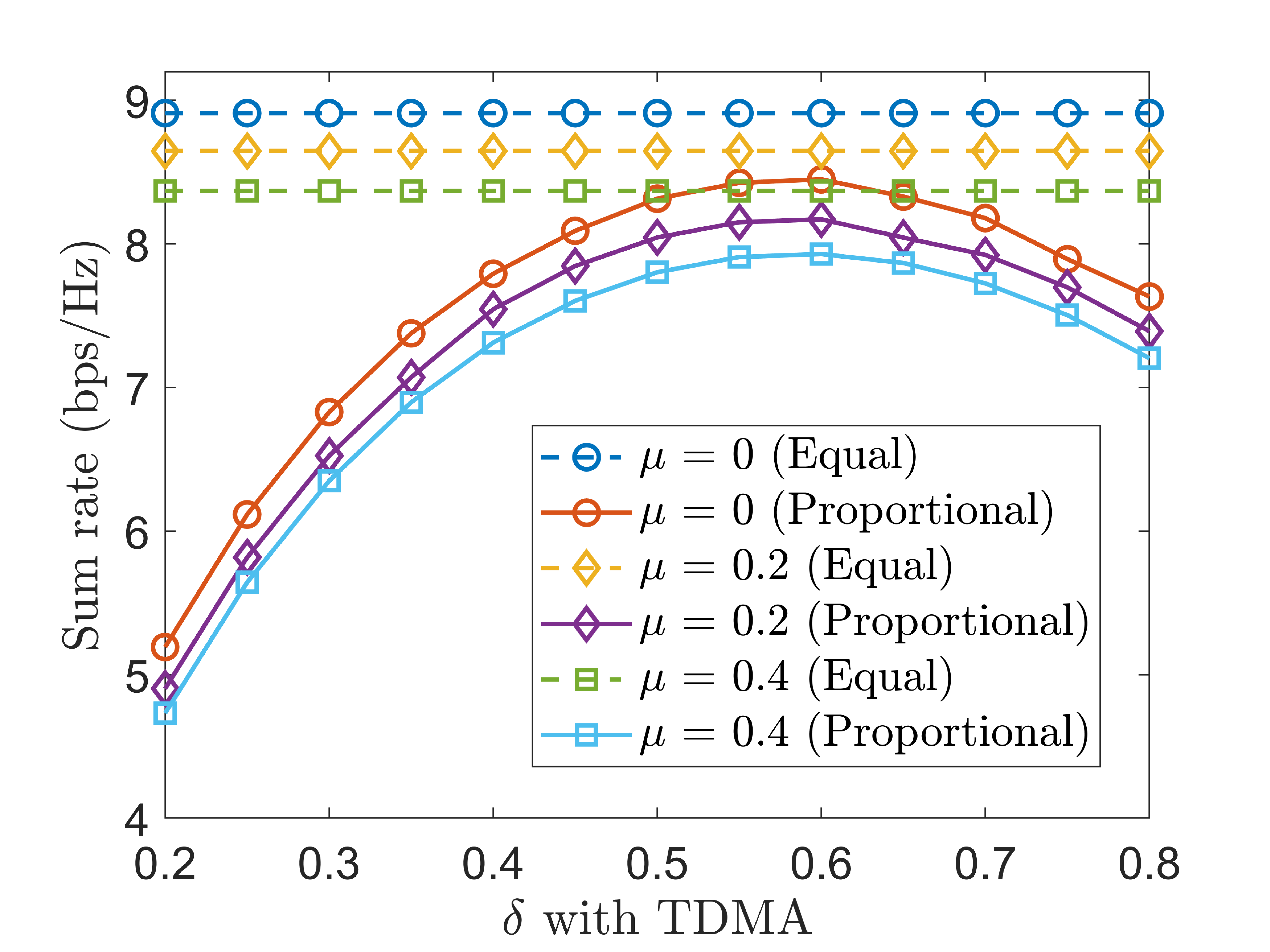}
		\subcaption{\centering TDMA.}
		\label{Na}
	\end{subfigure}
	\hfill
	\begin{subfigure}{0.49\linewidth}
		\centering
		\includegraphics[width=\linewidth]{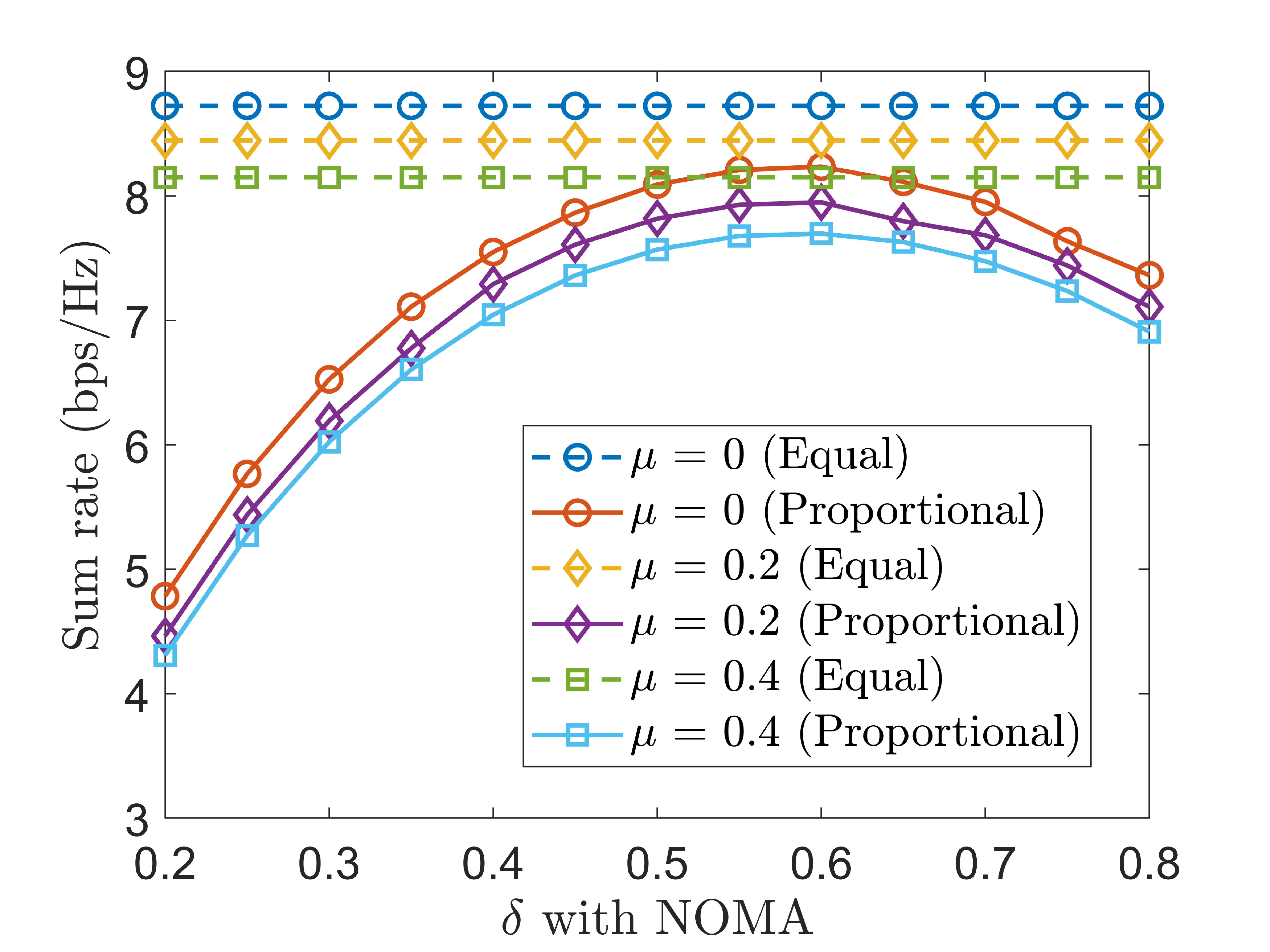}
		\subcaption{\centering NOMA.}
		\label{Nb}
	\end{subfigure}
	\caption{Sum rate versus $\delta$ with TDMA and NOMA protocols.}
	\label{delta}
\end{figure} 
Fig. {\ref{N}} illustrates the variation in the sum rate with respect to different numbers of PAs under $p_{c,k}=0~{\rm W}$ and $p_{c,k}=10^{-7}~{\rm W}$, respectively. As the number of PAs increases, the system can obtain superior beamforming gains to improve overall performance. We observe that the SPDE algorithm outperforms the EW algorithm when $N<4$. This is because smaller values of $N$ allows the SPDE algorithm to more effectively explore superior solutions, while the EW algorithm's performance is constrained by its discrete solution set. Moreover, the proposed system achieves a significantly higher sum rate compared to conventional antenna systems. {Although RIS provides additional reflecting paths for WPCN, its performance exceeds that of fixed-antenna WPCN but is still inferior to PA-WPCN due to the dual fading effect.} It is worth noting that when $p_{c,k}=0~{\rm W}$, the system achieves an identical sum rate under both TDMA and NOMA protocols, as previously demonstrated in the preceding section. 

Fig. {\ref{delta}} illustrates the impact of different power distribution factors 
$\delta$ on the sum rate of the considered system. Based on $\sum\nolimits_{i = 1}^I {{\delta ^2}{{\left( {1 + {\delta ^2}} \right)}^{i - 1}}}  \le \sum\nolimits_{j = 1}^J {{\delta ^2}{{\left( {1 + {\delta ^2}} \right)}^{j - 1}}} < 1, I<J$, we infer that the signal power within the waveguide cannot be fully allocated under the proportional power model. It can be observed that a larger value of $\delta$ allows more signal power to be radiated from the waveguide. However, it also results in allocating more power to the first PA, thereby weakening beamforming performance. Consequently, the selection of $\delta$ requires a trade-off. Simulation results demonstrate that setting $\delta$ to approximately $0.55$$-$$0.6$ yields the optimal system performance. {The optimal $\delta$ is observed to remain within the range of $0.55$$-$$0.6$ across various system configurations (e.g., varying $K$, $D_w$, etc.).} {In contrast, the ideal equal power model assumes the entire power budget is uniformly radiated, thus avoiding power residue and achieving higher rate gain.}

\section{Conclusion}
In this letter, a PA-WPCN under TDMA and NOMA protocols was investigated, leveraging the reconfigurable propagation capability of the PASS to enhance the downlink WPT and the uplink WIT. We formulated sum rate maximization problems under both protocols to solve the joint optimization of time slot allocation, devices' transmit power, and the positions of the PAs. To address the intractable original problems, we decoupled them as resource allocation and PA position optimization subproblems. Specifically, PAs position optimization subproblem was addressed by utilizing the proposed EW and SPDE algorithms. Simulation results demonstrate that the proposed PA-WPCN outperforms conventional fixed-antenna systems and reveal that the optimal PA power allocation ratio lies approximately in the range of $0.55$$-$$0.6$. 

\bibliography{Reference}
\bibliographystyle{IEEEtran}

\end{document}